\documentclass[letterpaper]{article} 
\usepackage{aaai25}  
\usepackage{times}  
\usepackage{helvet}  
\usepackage{courier}  
\usepackage[hyphens]{url}  
\usepackage{graphicx} 
\usepackage{natbib}  
\usepackage{caption} 
\usepackage{algorithm}
\usepackage{algorithmic}
\usepackage{amsmath}
\usepackage{amsthm}
\usepackage{booktabs}
\usepackage{multirow}
\usepackage[switch]{lineno}
\usepackage{siunitx,diagbox}

\usepackage{amsfonts}
\usepackage{url}

\usepackage{tikz}

\usetikzlibrary{positioning,fit}
\usepackage{pgfplots}
\pgfplotsset{compat=newest}
\usepgfplotslibrary{groupplots}

\usepgfplotslibrary{colorbrewer}
\usetikzlibrary{pgfplots.statistics,positioning,fit,pgfplots.colorbrewer}

\usepackage{caption}
\usepackage{subcaption}

\usepackage{colortbl}

\usepackage{forest}

\usepackage{xspace}
\usepackage{newfloat}
\usepackage{listings}
\DeclareCaptionStyle{ruled}{labelfont=normalfont,labelsep=colon,strut=off} 
\floatstyle{ruled}
\newfloat{listing}{tb}{lst}{}
\floatname{listing}{Listing}

\usepackage{niravstyle}

\newtheorem{definition}{Definition}
\newcommand{\method}{RAWL${\cdot}$E\xspace}

\newcommand{\nsahide}[1]{}
\newcommand{\BibTeX}{B\kern-.05em{\sc i\kern-.025em b}\kern-.08em\TeX}
\usepackage{soul}

\newcommand{\revisionst}[1]{\textcolor{red}{\st{#1}}}
\renewcommand{\revisionst}[1]{}

\newcommand{\NM}{\msf{NM}}
\newcommand{\EM}{\msf{EM}}
\newcommand{\DM}{\msf{DM}}

\title{Operationalising Rawlsian Ethics for Fairness in Norm-Learning Agents}
\author {
    Jessica Woodgate,
    Paul Marshall,
    Nirav Ajmeri
}
\affiliations {
    School of Computer Science\\
    University of Bristol\\
    Bristol BS8 1UB, UK\\
    jessica.woodgate@bristol.ac.uk, p.marshall@bristol.ac.uk, nirav.ajmeri@bristol.ac.uk
}

\begin{document}

\maketitle

\begin{abstract}
Social norms are standards of behaviour common in a society. 
However, when agents make decisions without considering how others are impacted, norms can emerge that lead to the subjugation of certain agents.
We present \method, a method to create ethical norm-learning agents.
\method agents operationalise maximin, a fairness principle from Rawlsian ethics, in their decision-making processes to promote ethical norms by balancing societal well-being with individual goals. 
We evaluate \method agents in simulated harvesting scenarios. 
We find that norms emerging in \method agent societies enhance social welfare, fairness, and robustness, and yield higher minimum experience compared to those that emerge in agent societies that do not implement Rawlsian ethics.
\end{abstract}

%

\section{Introduction}
\label{sec:introduction}

Social norms are standards of expected behaviour that govern a multi-agent system (MAS) and enable coordination between agents \cite{Levy+Griffiths2021ConventionEmergence,vonWright1963NormAction}. Norms can be established through top-down prescriptions or emerge bottom-up via interactions between agents \cite{Morris+19:norm-emergence}.
However, when agents are solely self-interested, norms may emerge that exploit some agents for the benefit of others.
Where ethics involves one agent's concern for another \cite{Murukannaiah+Singh-IC20-Ethics}, norms which result in the subjugation of agents are unethical.
If agents learn norms by appealing to existing behaviours in a society without evaluating how ethical those behaviours are, they risk perpetuating unethical norms.

Previous works on promoting norms which are considerate of others, such as Tzeng \etal \shortcite{Tzeng+COINE22-Fleur} and Dell'Anna \etal \shortcite{DellAnna-JAAMAS20-Runtime}, appeal to individual or societal preferences over values. 
Other works observe the behaviour of others to encourage cooperation:
Oldenburg and Zhi \shortcite{Oldenburg+Zhi-Xuan2024BayesianNorms} infer norms by observing apparent violations of self-interest;
Guo \etal \shortcite{Guo2020ContextNorm} 
learn contextual priority of norms from observing experts; 
Chen \etal \shortcite{Chen2017sociallyAware} imply norms through reciprocity.

However, approaches which appeal to preferences or existing behaviours to promote cooperation define ethical behaviour by reference to descriptive statements, which are statements that express what states of affairs are like
\cite{Kim2021takingPrinciples}. 
Attributing ethics descriptively may lead to the issue of deriving an \fsl{ought} from an \fsl{is}---just because something is the case doesn't mean it ought to be.
Where existing norms or behaviours are unethical, approaches that encourage cooperation through descriptive facts thereby risk propagating unethical norms which reflect \fsl{what is the case}, rather than \fsl{what ought to be}.

To mitigate the \fsl{is-ought} gap, we turn to 
\fsl{normative ethics}. Normative ethics is the study of practical means to determine the ethical acceptability of different courses of action \cite{Woodgate+Ajmeri-CSUR2024-Principles}. 
Normative ethics principles are justified by reason in philosophical theory.
These principles are normative in that they are prescriptive, indicating how things ought to be, rather than descriptive, indicating how things are \cite{Kim2021takingPrinciples}.

The principle of \fsl{maximin}---to maximise the minimum experience---is a widely respected fairness principle in normative ethics advanced by Rawls \shortcite{Rawls1958justiceAsFairness}.
Rawls states that in a society with unequal distribution which is not to the benefit of all, those benefiting the least should be prioritised.
We hypothesise that creating agents that promote the emergence of ethical norms, while avoiding the \fsl{is-ought} gap, is aided by an appeal to Rawlsian ethics \cite{Woodgate+Ajmeri-AAMAS22-BlueSky}.

\paragraph{Contribution} We propose \method, a novel method to design socially intelligent norm-learning agents that consider others in individual decision making by operationalising the principle of maximin from Rawlsian ethics. A \method agent includes an ethics module that applies maximin to assess the effects of its behaviour on others. 

\paragraph{Novelty} Operationalising Rawlsian ethics in learning agents to enable explicit norm emergence is a novel contribution.
\method goes beyond existing works on norm-learning agents: 
Ajmeri \etal's \shortcite{Ajmeri-AAMAS20-Elessar} agents incorporate ethical decision-making, but do not incorporate learning.
Agrawal \etal \shortcite{Agrawal+IJCAI22-SIGA} address the emergence of explicit norms, but optimise norms based on the sum of payoffs received by other agents, which might be unfair for some agents.
Zimmer \etal \shortcite{zimmer2021fairPolicies} and
Balakrishnan \etal \shortcite{balakrishnan2022SCALES} operationalise Rawlsian ethics in learning agents, but do not consider the role of norms.
As a \method agent gains experience, it learns to achieve its goals whilst behaving in ways that support norms which prioritise those who are least advantaged in situations of unequal resource distribution.

We evaluate \method agents in two simulated harvesting scenarios implemented in reinforcement learning (RL).
We find that 
(1) \method agents learn ethical norms that promote the well-being of the least advantaged agents, and 
(2) \method agent societies yield higher social welfare, fairness, and robustness than agents who do not operationalise Rawlsian ethics in their individual decision-making.

\paragraph{Organisation}
\label{sec:organisation}

Section~\ref{sec:related} explores related works and gaps. 
Section~\ref{sec:method} describes our method. 
Section~\ref{sec:simulation} presents the simulation environment used to evaluate \method agents. 
Section~\ref{sec:results} discusses results of our simulation experiments. 
Section~\ref{sec:conclusion} concludes with a discussion of future directions.

\section{Related Works}
\label{sec:related}

Research on combining normative ethics with norm emergence and learning is relevant to our contributions.

\paragraph{Normative Ethics and Norm Emergence}
Ethical norm emergence has been examined through the lens of agent roles. Anavankot \etal \shortcite{Anavankot2023NormEntrepreneurship} propose norm entrepreneurs that influence the dynamics of norm-following behaviours and thus the emergence of norms.
Vinitsky \etal \shortcite{Vinitsky2023sanctions} study norm emergence through sanction classification.
Levy and Griffiths \shortcite{Levy+Griffiths2021ConventionEmergence} manipulate rewards using a central controller to enable norm emergence.
Neufeld \etal \shortcite{neufeld_enforcing_2022} use deontic logic to implement a normative supervisor module in RL agents.
Yaman \etal's \shortcite{Yaman_2023sanctioning} agents sanction one another to encourage effective divisions of labour.
Maranh{\~a}o \etal \shortcite{maranhao2022normativeChange} formally reason about normative change.
However, a gap remains in agents learning norms based on what ought to be the case, rather than what is.
We address this gap by implementing principles from normative ethics to encourage the emergence of norms that can be justified independently to a specific situation.

Traditional approaches encourage norm emergence by maximising social welfare---how much society as a whole gains. Shoham and Tennenholtz \shortcite{Shoham1997SocialConventions} promote highest cumulative reward. Yu \etal \shortcite{yu2014CollectiveLearning} utilise majority vote.
Agrawal \etal \shortcite{Agrawal+IJCAI22-SIGA} sum the payoffs for different stakeholders.
Focusing on social welfare alone may lead to situations where a minority is treated unfairly for the greater good \cite{Anderson2004machineEthics}, and mutual reward does not specify how to coordinate fairly \cite{Grupen2021CooperativeMF}.
To mitigate weaknesses associated with only maximising social welfare, we implement Rawlsian ethics, emphasising improving the minimum experience.

\paragraph{Normative Ethics and Learning}
Jing and Doorn \shortcite{jing2020confucianResponsibility} emphasise the importance of focusing on positive standards alongside preventative ethics, which involves negative rules denoting wrongdoing. As ethics is dynamic, it may not always be possible to determine which behaviours to restrict.

Svegliato \etal \shortcite{Svegliato_Nashed_Zilberstein_2021} implement divine command theory, prima facie duties, and virtue ethics; Nashed \etal \shortcite{Nashed2021MoralCommunities} implement the veil of ignorance, golden rule and act utilitarianism.
Dong \etal \shortcite{dong2024egoism} optimise federated policies under utilitarian and egalitarian criteria.
A gap exists, however, in applying normative ethics in RL to norm emergence. 
\method addresses that gap.

\section{Method}
\label{sec:method}
We now present our method to design \method agents who operationalise Rawlsian ethics to support the emergence of ethical norms.

\subsection{Schematic}
\label{sec:schematics}


\begin{definition}
Environment $\mathbb{E}$ is a tuple $\langle AG, D, \mathcal{N} \rangle$ where, 
    $AG=\{ag_1,...,ag_n\}$ is a set of agents; 
    $D$ is the amount of total resources; 
    $\mathcal{N}$ is the set of norms.
\end{definition}




\begin{definition}
A \fsl{\method agent} is a tuple $\colon$ $\langle d, \upsilon, G, A, Z, \NM, \EM, \DM \rangle$ where,
$d \in D$ is the amount of resources to which the agent has access; 
$\upsilon$ is a measure of its well-being;  
$G$ is the set of goals $g_1,...,g_l$; 
$A$ are the actions available to the agent to help achieve its goals; 
$Z$ are the behaviours which the agent has learned; 
$\NM$ is its norms module;
$\EM$ is its ethics module; and
$\DM$ is its interaction module.
\end{definition}

\begin{definition}
A \fsl{goal} $g \in G$ is a set of favourable states an agent aims to achieve. 
\end{definition}
        
\begin{definition}
A behaviour $\zeta \in Z$ is a tuple $\langle \msf{pre}, \msf{act}\rangle$, where
    $\msf{pre}\in \text{Expr}$ is its precondition;
    $\msf{act}\in \text{Expr}$ is its action; and 
    $\text{Expr}$ is any logical expression that can be evaluated as either true or false based on the values of its variables.  
\end{definition}

A behaviour has a precondition denoting the conditions within which the behaviour arises, and a postcondition, which is the action implied by the precondition. Each agent keeps a record of their learnt behaviours.

A behaviour is encoded in the form of an if-then rule:

\begin{lstlisting}[numbers=none]
    <behaviour> ::= IF <pre> THEN <act>
\end{lstlisting}

\begin{definition}
A norm $n \in \mathcal{N}$, where $\mathcal{N} \subseteq Z$, is a behaviour 
adopted by a society.
\end{definition}
Norms are the prescription and proscription of agent behaviour on a societal level \cite{savarimuthu_identifying_2013}. 

\begin{definition}
$\mathcal{N}$, where $\mathcal{N} \subseteq Z$, denotes the set of emerged norms, i.e., the behaviours adopted by the society as norms, which form a normative system describing a society.
\end{definition}
        
Norms emerge when the same behaviours are adopted by other agents \cite{Tuomela1995SocialNotations}. Norm emergence is accepted to have happened when a predetermined percentage of the population adopt the same behaviours. As following previous literature, we assume a norm to have emerged when it reaches 90\% convergence \cite{Kittock1995EmergentCA}.

\begin{definition}
A sanction $F$ represents a positive or negative reaction to behaviour which provides feedback to the learner in the form of a reward.
\end{definition}

Sanctions are positive or negative reactions to behaviour which help enforce norms. A self-directed sanction is a sanction directed towards and affecting only its sender \cite{Nardin-KER16-Classifying}. The self-directed sanction provides feedback to the learner as a reward.

\subsection{Interaction and Norm Learning}
\label{sec:agent-components}

To make decisions and pursue their goals, \method agents use ethics module, norms module, and interaction module.
\subsubsection{Ethics Module}
\label{sec:ethics-module}

Ethics module, $\EM$, assesses how actions affect the well-being of other agents.
%
To evaluate the well-being of others, \method agents implement Rawlsian ethics. 
Adapted from Leben \shortcite{Leben2020Normative}, an ethical utility function 
$u(d) \rightarrow (\upsilon)$ models a distribution of resources, where $d$ is a vector of resource distribution which sums to $D$, the amount of total resources, and $(\upsilon)$ is a measurement of well-being for agents considering that resource distribution. 
Where $w$ is a vector of inputs (e.g., observed well-being of agents), Rawlsian ethics is expressed as:

\begin{equation}
MA(d) = min_{w} \quad u(d,\upsilon_i)
\end{equation}

Via $MA(d)$, the ethics module evaluates whether the agent's action improves the minimum experience. 
It generates a positive self-directed sanction $\xi$ if an action improves the minimum experience, and a neutral or negative sanction $-\xi$ if it does not change or worsens. 
In analogy to the real world, a positive sanction represents happiness from helping others, while a negative sanction represents guilt.
To implement $MA$, ethics module takes as input $U_t$ and $U_{t+1}$, where $U$ is a vector of well-being $\upsilon_1,\ldots,\upsilon_n$ for all agents $ag_1,\ldots,ag_n$ at times $t$ and $t+1$. 
Ethics module identifies the minimum experience $min_{w}u(d,\upsilon)$ at $t$ and $t+1$, storing the results in $\upsilon\fsub{min}_t$ and $\upsilon\fsub{min}_{t+1}$, respectively.
Therefore:

\begin{equation}
    F_{t+1}(s_t, s_{t+1})= 
\begin{cases}
    \xi,& \text{if } \upsilon_{\text{min}_t} < \upsilon_{\text{min}_{t+1}}\\
     0,              &  \quad \upsilon_{\text{min}_t} = \upsilon_{\text{min}_{t+1}}\\
    - \xi   &  \quad \upsilon_{\text{min}_t} > \upsilon_{\text{min}_{t+1}}
\end{cases}
\end{equation}

Algorithm~\ref{alg:ethics-module} describes internals of the ethics module. 
The inputs are $U_t$ and $U_{t+1}$. 
To implement $MA$, store $\upsilon_{\text{min}_t}$ and $\upsilon_{\text{min}_{t+1}}$ (lines 1--2). 
Compare $\upsilon_{\text{min}_t}$ and $\upsilon_{\text{min}_{t+1}}$ to assess how action $a$ taken in $s_t$ affected $\upsilon_{\text{min}_{t+1}}$ (Line 3). 
Generate sanction $F_{t+1}$ (Lines 4--7). 
Output $F_{t+1}$ for interaction model to combine with environmental reward $r_{t+1}$ through reward shaping so that $r'_{t+1} = r_{t+1} + F_{t+1}$. (Line 8).

\begin{algorithm}[!htb]
    \caption{Ethics module.}
    \label{alg:ethics-module}
    \textbf{Input}: $U_t$, $U_{t+1}$\\
    \textbf{Output}: $F_{t+1}$
    \begin{algorithmic}[1] 
        \STATE $\upsilon_{\text{min}_t} \gets$ getMinExperience($U_t$)
        \STATE $\upsilon_{\text{min}_{t+1}} \gets$ getMinExperience($U_{t+1}$)
        \IF{$\upsilon_{\text{min}_{t+1}} > \upsilon_{\text{min}_{t}}$}
            \STATE $F_{t+1} = \xi$
        \ELSIF{$\upsilon_{\text{min}_{t+1}} == \upsilon_{\text{min}_{t}}$}
            \STATE $F_{t+1} = 0$ 
        \ELSE
            \STATE $F_{t+1} = -\xi$
        \ENDIF
        \STATE \textbf{return} $F_{t+1}$
    \end{algorithmic}
\end{algorithm}

\subsubsection{Norms Module}
\label{sec:norms-module}

Norms module, $\NM$, tracks patterns of behaviour the agent learns. 
Norms module stores behaviours in a behaviour base and norms in a norm base.
For each behaviour, it computes and stores the numerosity $\mathrm{num}$, obtained from the number of times the behaviour is used, and the reward $r'$ (described in interaction module) received from using the behaviour. 
The fitness of each behaviour $\tau$ is obtained from $\mathrm{num} \cdot r'$ decayed over time.
Where $\eta$ is the age of the behaviour and $\lambda$ is the decay rate, 
\begin{equation}\label{equ:norms-fitness}
    \tau(\zeta) = \mathrm{num} \cdot r' \cdot \lambda^{\eta}
\end{equation}

Algorithm~\ref{alg:norms-module} describes the internals of the norm module. Inputs to the norm module include $\nu_t, a_t, r'_{t+1}$, where $\nu_t$ is the precondition obtained from the agent's view of state $s_t$ (for scalability, $\nu_t$ is a subset of $s_t$); $a_t$ is the action taken in $s_t$. Norms module searches the behaviour base to retrieve a behaviour matching $\langle \msf{pre}, \msf{act}\rangle$ to $\nu_t, a_t$ (line 1). 
If there is a matching behaviour, update $\tau(\zeta)$ (lines 2--3). 
If there is no match, behaviour learner creates a new behaviour with $\nu_t, a_t$ and adds it to behaviour base (lines 5--6). 
Every $t\fsub{clip\_behaviours}$ steps, if behaviour base exceeds the maximum capacity, behaviour base is clipped to the maximum capacity by removing the least fit behaviours (lines 8--9). 
Norms module compares behaviour base with norm base shared by the society and stores emerged norms in norm base (line 10).

\begin{algorithm}[!htb]
    \caption{Norms module.}
    \label{alg:norms-module}
    \textbf{Input}: $\nu_t, a_t$
    \begin{algorithmic}[1] 
        \STATE $\zeta \gets$ behaviourBase.retrieve($\nu_t, a_t$)
        \IF{$\zeta !=$ None}
            \STATE behaviourBase.updateFitness($\zeta$)
        \ELSE
            \STATE $\zeta \gets$ behaviourLearner.create($\nu_t$, $a_t$)
            \STATE behaviourBase.add($\zeta$)
        \ENDIF
        \IF{t $\%$ clipNorm is 0 and len(behaviourBase) $>$ maxLen}
            \STATE behaviourBase.clip()
        \ENDIF
        \STATE normBase.updateEmergedNorms(behaviourBase)
    \end{algorithmic}
\end{algorithm}

\subsubsection{Interaction Module}
\label{sec:decision-making-module}

Interaction module, $\DM$, implements RL with deep~Q~network (DQN) architecture \cite{sutton_reinforcement_2018}. Via DQN, \method agent learns a behaviour policy to achieve goals while promoting ethical norms. At each time step $t$, agent selects a batch of $B$ random experiences to train its Q network against its target network, computing the Huber loss \cite{Huber1964RobustEO}. To prevent overfitting, every $C$ steps weights of target network are updated to weights of the Q network $\theta$. At each step, agent receives an observation of the environment, a vector of features $x(s)$ visible in state $s$, which it stores in the experience replay buffer. Each feature of $x(s)$ coresponds to a feature in the agent's DQN. With probability $\epsilon$, agent selects an action randomly or using DQN. Using DQN, actions $a \in A$ are selected which policy $\pi(s)$ estimates will maximise expected return and help achieve goals $G$. Agent acts asynchronously and receives a reward from its environment $r$. $\DM$ obtains shaped reward $F_{t+1}$ from $\EM$. To encourage an agent to learn behaviours which promote ethical norms whilst pursuing goals, $\DM$ combines self-directed sanction $F_{t+1}$ with environmental reward $r_{t+1}$ through reward shaping so that $r'_{t+1} = r_{t+1} + F_{t+1}$. Transition $(a_t, s_t, s_{t+1}, r'_{t+1})$ is stored in experience replay buffer. $\DM$ obtains view $\nu_t$ from state $s_t$ and passes $\nu_t$ to $\NM$ for norm learning.

Algorithm~\ref{alg:decision-module} outlines the interaction module. Input environmental observation at $s_t$, which includes environment state, agent's resources $d$, and well-being $\upsilon_1,\ldots,\upsilon_n$ of all agents $ag_1,\ldots,ag_n$. 
Deterministic policy $\Pi(\theta,a)$ defines the agent's behaviour in $s_t$ to output action $a_t$ (Line~1). 
After acting, observe $r_{t+1}$, $s_{t+1}$ (Line~2); obtain well-being vectors $U_t$ and $U_{t+1}$ with $\upsilon_1,\ldots,\upsilon_n$ obtained from $s_t$ and $s_{t+1}$ (Lines~3--4); pass $U_t$ and $U_{t+1}$ to $\EM$ to obtain $F_{t+1}$ (Line 5); obtain $r'_{t+1}$ from $r_{t+1}$ and $F_{t+1}$ (Line~6); update $\Pi(\theta,a)$ (Line~7); obtain $\nu_t$ from $s_t$ (Line 8); pass $\nu_t$ to $\NM$ to learn and store behaviours and norms (Line~9). 

\begin{algorithm}[!htb]
    \caption{Interaction module.}
    \label{alg:decision-module}
    \textbf{Input}: $s_t$
    \begin{algorithmic}[1] 
        \STATE $a_t \gets \pi(s_t)$ \COMMENT{\textit{Obtain action from policy}}
        \STATE $r_{t+1}$, $s_{t+1} \gets$ act($a_t$) \COMMENT{\textit{Perform action, observe $r_{t+1}, s_{t_+1}$}}
        \STATE $U_t \gets $ getWellbeing($s_t$) \COMMENT{\textit{Obtain well-being}}
        \STATE $U_{t+1} \gets $ getWellbeing($s_{t+1}$) 
        \STATE $F_{t+1} \gets$ EthicsModule($U_t, U_{t+1}$) \COMMENT{\textit{Obtain sanction}}
        \STATE $r'_{t+1} \gets r_{t+1} + F_{t+1}$ \COMMENT{\textit{Shape reward}}
        \STATE $\Pi(\theta,a) \gets$ update($\Pi$, $s_t$, $r'_{t+1}$, $s_{t+1}$) \COMMENT{\textit{Update policy}}
        \STATE $\nu_t \gets$ getView($s_t$) \COMMENT{\textit{Obtain view of $s_t$}}
        \STATE NormsModule($\nu_t, a_t, r'_{t+1}$) \COMMENT{\textit{Update norms module}}
    \end{algorithmic}
\end{algorithm}

\section{Simulation Environment}
\label{sec:simulation}

\begin{figure*}[!htb]
    \centering
    \begin{subfigure}[t]{.42\textwidth}
    \centering
    \includegraphics[width=\columnwidth]{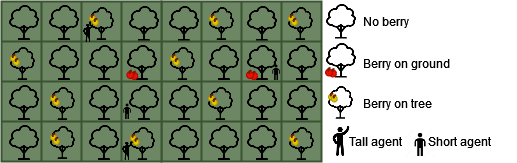}
    \caption{Capabilities harvest. Agents move freely through the grid but can only harvest certain berries. Some berries are on the ground, only visible by short agents. Others are in trees, only visible by tall agents. Agents can learn to throw berries to one another across the grid.}
    \end{subfigure} 
    \hfill
    \begin{subfigure}[t]{.56\textwidth}
    \centering
    \includegraphics[width=\columnwidth]{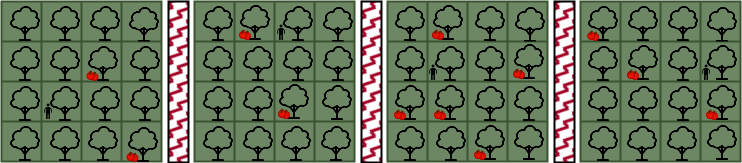}
    \caption{Allotment harvest. Agents are assigned a certain allotment in a community garden. Agents can only harvest berries within their allotment. Each allotment has a different amount of berries that grow there. Agents can learn to throw berries to agents in other allotments.}
    \end{subfigure} 
    
    \caption{Harvesting environment.  
    (a) Capabilities harvest scenario explores how agents learn to identify and reach desired berries while considering the well-being of the society.
    (b) Allotment harvest scenario explores how agents learn to harvest within their desired areas while considering the well-being in the society.}
    \label{fig:berry-world}
\end{figure*}

We evaluate \method agents in a simulated harvesting scenario where they forage for berries. Cooperative behaviours may emerge, such as agents learning to throw berries to one another. To demonstrate the efficacy of modular ethical analysis, the scenario includes environmental rewards for cooperation. Figure~\ref{fig:berry-world} shows our harvesting environment.

\subsection{Scenario}
\label{sec:scenario}

The environment represents a cooperative multi-agent scenario with a finite population of agents on a $o \times p$ grid. Time is represented in steps. At the beginning of each episode, the grid is initialised with $k=4$ agents, and $b\fsub{initial}=12$ berries at random locations. An agent begins with $h\fsub{initial}=5.0$ health. Agents may collect berries, throw berries to other agents, or eat berries. An agent receives a gain in health $h\fsub{gain}=0.1$ when it eats a berry. Agent health decays $h\fsub{decay}=-0.01$ at every time step. An agent dies if its health level reaches 0 and episode ends when all agents have died. 
 Appendix~\ref{appx:hyperparameter-selection}, Tables~\ref{tbl:simulation-parameter-selection} and ~\ref{tbl:decision-parameter-selection} provide complete list of simulation parameters and parameters for the interaction module.



Agents act asynchronously, in a different random order on each step of the simulation. At each step, each agent $ag_i$ decides to move (north, east, south, west), eat a berry, or throw a berry to another agent $ag_j$ if $ag_i$ has at least $h\fsub{throw}=0.6$ health. When an agent has eaten a berry, a berry regrows at a random location on the grid. At each step, an agent forages for a berry in its location. An agent observes its health, its berries, distance to the nearest berry, and each agent's well-being. Well-being is represented by a function of an agent's health and number of berries it has in its bag:
\begin{equation}\label{equ:days-left}
    ag\fsub{well-being} = \frac{ag\fsub{health} + (ag\fsub{berries} \times h\fsub{gain})} {h\fsub{decay}}
\end{equation}

For each agent, at each time step:
\begin{enumerate}[nosep,label=(\arabic*)]
    \item Receive observation $s_t$
    \item Choose $a$ using DQN: move (north, south, east, west), eat, throw
    \item Forage for berry; update health ($h\fsub{decay}$ at each step, $h\fsub{gain}$ if berry eaten)
    \item Receive transition: $r_{t+1}$, $s_{t+1}$, check if done
    \item Pass transition to Q network to learn
    \item Every $C$ steps, update $\theta$ of target network
    \item Pass transition to norms module, update norm base
    \item Check health, if agent has died remove from the grid
\end{enumerate}

For testing, we run each simulation $e=2000$ times, with each simulation running until all agents have died, or a maximum of $t\fsub{max}=50$ steps. We select these numbers empirically.
Agents clip behaviour every $t\fsub{clip\_behaviours}$ steps, clip norm base every $t\fsub{clip\_norms}$ steps, and check for emerged norms every step. Table~\ref{tbl:norm-parameters} lists the norm parameters.

\begin{table}[!htb]
    \centering
    \caption{Norm parameters.}
    \label{tbl:norm-parameters}
    \begin{tabular}{l p{4.5cm}  S}
        \toprule 
         {Parameter} & {Description} & {Value}
        \\\midrule
        $t\fsub{clip\_behaviours}$ & Clip behaviour base frequency & 10.0
        \\
        $t\fsub{clip\_norms}$ & Clip norm base frequency & 5.0
        \\\bottomrule
    \end{tabular}
\end{table}

\subsection{Society Types for Evaluation}
\label{sec:societies}

We implement two types of agent societies for evaluation.
\begin{description}[nosep,leftmargin=0em]
    \item[Baseline Cooperative: DQN] A society consists of standard DQN agents who do not implement an ethics module but receive environmental rewards for cooperative behaviour. 
    DQN agent makes decisions according to its observations and expected reward.
    \item[\method: Rawlsian DQN] A society of \method agents act in ways that promote Rawlsian ethics. \method agent makes decisions according to its observations and expected reward, considering the well-being of all agents.
\end{description}

\subsection{Environmental Rewards}
\label{sec:actions}

An agent receives a positive reward if it forages for a berry in a location where a berry is growing, if it eats a berry when it has berries in its bag, or if it survives to the end of the episode.
An agent receives a negative reward if it attempts to eat or throw a berry to another agent when it doesn't have any, or if it dies. Agent deaths are included in raw rewards to provide incentives for societies to survive.

Self-directed sanction of a \method agent is 0.4 if the minimum experience was improved, --0.4 if the agent could have improved the minimum experience but did not (i.e., if an action was available to improve the minimum experience but the agent chose an alternative action), and 0 otherwise.

To avoid obvious results by giving \method agents additional rewards, we normalise rewards between baseline and \method agents such that \method agents receive lower raw rewards. This allows for fairer comparison between societies. Table~\ref{tbl:rewards} summarises rewards an agent receives; in Appendix~\ref{appx:rewards}, Table~\ref{tbl:complete-rewards} displays the complete list of rewards.

\begin{table}[!htb]
    \centering
    \caption{Rewards received by an agent. To avoid obvious results by giving \method agents more rewards, we normalise rewards between baseline and \method agents.}
    \label{tbl:rewards}
    \begin{tabular}{@{}p{5.5cm}@{} SS@{}}
        \toprule 
         {Action} & {Baseline} & {\method}
        \\\midrule
        Survive episode & 1.0 & 1.0
        \\
        Eat berry & 1.0 & 0.8
        \\
        Forage where berry is & 1.0 & 0.8
        \\
        Throw berry to others & 0.5 & 0.5
        \\
        Die & -1.0 & -1.0
        \\
        Improve minimum experience & 0.0 & 0.4
        \\
        Did not improve minimum experience & 0.0 & -0.4
        \\\bottomrule
    \end{tabular}
\end{table}

\subsection{Metrics and Hypotheses}
\label{sec:metrics+hypotheses}

Emerged norms $\mathcal{N}$ describe the standards of expected behaviour in a society. 
To evaluate $\mathcal{N}$, we examine cooperative norms which emerge by their fitness and numerosity. We assess the effects of those norms on societal outcomes with the following metrics and hypotheses.

\subsubsection{Variables}
\label{sec:variables}
To quantitatively assess societal outcomes, for each simulation run, we record the following variables:

\begin{description}[nosep,leftmargin=0em]
    \item[V\fsub1 ($ag\fsub{well-being}$)] Number of days an agent has left to live, a function of number of berries an agent carries and their current health (Equation~\ref{equ:days-left}).
    \item[V\fsub2 ($ag\fsub{resource}$)] Number of berries eaten by an agent. 
    
\end{description}


\subsubsection{Metrics}
\label{sec:metrics}
To assess fairness on an individual and at societal level, we compute the metrics M\fsub1~(inequality) and M\fsub2~(minimum experience) on each variable. 

\begin{description}[nosep,leftmargin=0em]
    \item[M\fsub1 (inequality)] Gini index across the society. Lower is better. 0 denotes perfect equality; 1 denotes perfect inequality.
        
    \item[M\fsub2 (minimum experience)] Lowest individual experience across the society. Higher is better. 
\end{description}

To assess the sustainability of the society, we compute the metrics M\fsub3~(social welfare) and M\fsub4~(robustness).
\begin{description}[nosep,leftmargin=0em]
    \item[M\fsub3 (social welfare)] How much society as a whole gains \cite{Mashayekhi+TAAS22-Cha}. Higher is better.
    
    \item[M\fsub4 (robustness)] Length of episode. 
    Higher is better.
\end{description}

Appendix~\ref{appx:metrics} includes further description of the metrics.

\subsubsection{Hypotheses}
\label{sec:hypotheses}

We evaluate the following hypotheses. Null hypotheses for each indicate no difference.

\begin{description}[nosep,leftmargin=0em]
    \item[H\fsub1 (minimum experience)] Norms emerging in \method society lead to higher minimum individual experience. 
    
    \item[H\fsub2 (inequality)] Norms emerging in \method society lead to lower inequality. 
    
    \item[H\fsub3 (social welfare)]
    Norms emerging in \method society lead to higher social welfare. 
    
    \item [H\fsub4 (robustness)] Norms emerging in \method society lead to higher robustness. 
\end{description}

For each hypotheses, we test the significance and compute effect size. 
For significance, we conduct Mann-Whitney U test which is a non-parametric test for comparing two independent groups \cite{MannWhitneyU1947}. We use Mann-Whitney U because the sample size $k$ is small. $p<0.01$ indicates significance. 
For effect size, we compute Cohen's d which assesses the magnitude of difference between means, standardised by the pooled standard deviation \cite{Cohen-88-Statistics}, calculated as $\frac{\bar{x}_1 - \bar{x}_2}{s_{pooled}}$, where $<$0.2 (negligible), [0.2,0.5) (small), [0.5,0.8) (medium), and $\geq$0.8 (large).

\section{Experimental Results}
\label{sec:results}

To evaluate the behaviour of \method agents, we run agents in two experiment scenarios with different demonstrations of unequal resource allocation. For testing, we run $e=2000$ episodes, with each episode running until $t\fsub{max}=50$, or until the agents die. 
For qualitative analysis, we examine the emerged norms and actions promoted. For quantitative analysis, we examine fairness and sustainability metrics.


\subsection{Emerged Norms}

\method agent's norms model learns emerging norms from patterns of behaviour. 
To evaluate these norms, we run $e$ episodes for each society and store $\mathcal{N}$ from each episode. 
At each step, agents compare behaviour bases and store norms repeated by 90\% of agents in shared norm base $\mathcal{N}$. 

We observe that in both harvest scenarios, \method agents learn more cooperative norms of throwing berries than the baseline society, such as:

\begin{lstlisting}[numbers=none]
IF <high health, medium berries, low neighbour well-being> THEN <throw>
\end{lstlisting}

To evaluate $\mathcal{N}$ over $e$ episodes, we examine the numerosity $\mathrm{num}$ obtained from the times the norm is used, and fitness $\tau$ (Equation~\ref{equ:norms-fitness}) of cooperative norms. We find that \method agents learn cooperative norms with higher fitness and use cooperative norms more, indicated by higher numerosity. 
Appendix~\ref{appx:cooperative-norms}, Table~\ref{tbl:norms-summary} provides additional details of emerged cooperative norms. 
Appendix~\ref{appx:emerged-norms} provides the complete list of emerged norms.


\subsection{Simulation}

To quantitatively assess how ethical the normative system is, we analyse fairness and sustainability metrics of social welfare, inequality, minimum experience, and robustness. Table~\ref{tbl:results-summary} summarises results for the allotment harvest; Appendix~\ref{appx:simulation-results} includes additional results.
We find that the results are consistent across both scenarios with method agent societies having higher social welfare, lower inequality, higher minimum experience, and higher robustness.

\begin{table}[!htb]
    \sisetup{round-mode=places,
             round-precision=2,
             table-column-width=3.5em 
             }
    \centering
    \caption{Comparing $ag\fsub{resource}$, inequality, minimum experience, and robustness of baseline and \method societies in allotment harvest scenario. Grey highlight indicates best results with significance at $p<0.01$. 
    }
    \label{tbl:results-summary}
    
    \begin{tabular}{@{}l@{~}p{1.3cm}p{1cm}SS@{~~}S@{}}
        \toprule
                    & \multirow{2}{*}{Metrics} & \multirow{2}{*}{Variable} & \multicolumn{2}{c}{Mean $\bar{x}$} 
                    & \multicolumn{1}{c}{\multirow{2}{*}{Cohen's d}} \\
                    \cmidrule(lr){4-5}
                    & & & {Baseline} & {\method} &  \\\midrule
       \multirow{2}{*}{M\fsub1} & \multirow{2}{*}{Inequality} 
       & {$ag\fsub{well-being}$} & 0.2 & \cellcolor{gray!20}0.1 & 1.58\revisionst{0.6} \\
       & & {$ag\fsub{resource}$} & 0.14 & \cellcolor{gray!20}0.06 & 1.32\revisionst{0.73} \\\cmidrule{1-6}
       \multirow{2}{*}{M\fsub2} & \multirow{2}{*}{\parbox{1.8cm}{Minimum experience}}
       & {$ag\fsub{well-being}$} & 7.18 & \cellcolor{gray!20}10.82 & 3.09\revisionst{0.85} \\ 
       & & {$ag\fsub{resource}$} & 3.79 & 4.5 & 0.27\revisionst{0.13} \\\cmidrule{1-6}
       \multirow{2}{*}{M\fsub3} & \multirow{2}{*}{\parbox{1.5cm}{Social welfare}}
       & {$ag\fsub{well-being}$} & 51.5 & \cellcolor{gray!20}59.8 & 0.64\revisionst{0.29} \\
       & & {$ag\fsub{resource}$} & 18.94 & 20.6 & 0.14\revisionst{0.07} \\\cmidrule{1-6}
       M\fsub4 & {Robustness} & & 47.36 & \cellcolor{gray!20}48.19 & 0.11\revisionst{0.089} \\ \bottomrule       
    \end{tabular}
\end{table}

\paragraph{H\fsub1 (inequality)}

\begin{figure}[!htb]
\begin{subfigure}[t]{.48\columnwidth}
    \begin{tikzpicture}
      \begin{axis}[
          height=1.05\columnwidth,
          width=0.95\columnwidth,
          ylabel={Gini $ag\fsub{well-being}$},
          ymin=0,
          ymax=0.3,
          xtick={1,2},
          xticklabels={Baseline, \method},
          boxplot/draw direction=y,
          boxplot/box extend=0.3,
      ]
      
      \addplot+[boxplot, mark=*, mark size=1,] table[y=baseline, col sep=comma] {data/harvest_env/abilities/gini_days_left_to_live.csv};
      \addplot+[boxplot, mark=*, mark size=1,] table[y=maximin, col sep=comma] {data/harvest_env/abilities/gini_days_left_to_live.csv};
    
      \end{axis}
    \end{tikzpicture}
    \caption{Capabilities harvest. 
    }
    \label{fig:gini-well-being-abilities}   
\end{subfigure}
\hfill
\begin{subfigure}[t]{.48\columnwidth}
    \begin{tikzpicture}
      \begin{axis}[
          height=1.05\columnwidth,
          width=0.95\columnwidth,     
          ymin=0,
          ymax=0.3,          
          xtick={1,2},
          xticklabels={Baseline, \method},
          boxplot/draw direction=y,
          boxplot/box extend=0.3,
      ]
      
      \addplot+[boxplot, mark=*, mark size=1,] table[y=baseline, col sep=comma] {data/harvest_env/allotment/gini_days_left_to_live.csv};
      \addplot+[boxplot, mark=*, mark size=1,] table[y=maximin, col sep=comma] {data/harvest_env/allotment/gini_days_left_to_live.csv};
    
      \end{axis}
    \end{tikzpicture}
    \caption{Allotment harvest.
    }
    \label{fig:gini-well-being-allotment}
\end{subfigure}


\begin{subfigure}[t]{.48\columnwidth}
    \begin{tikzpicture}
      \begin{axis}[
          height=1.05\columnwidth,
          width=0.95\columnwidth,
          ylabel={Gini $ag\fsub{resource}$},
          ymin=0,
          ymax=0.4,
          xtick={1,2},
          xticklabels={Baseline, \method},
          boxplot/draw direction=y,
          boxplot/box extend=0.3,
      ]
      
      \addplot+[boxplot, mark=*, mark size=1,] table[y=baseline, col sep=comma] {data/harvest_env/abilities/gini_berries_consumed.csv};
      \addplot+[boxplot, mark=*, mark size=1,] table[y=maximin, col sep=comma] {data/harvest_env/abilities/gini_berries_consumed.csv};
    
      \end{axis}
    \end{tikzpicture}
    \caption{Capabilities harvest. 
    }
    \label{fig:gini-berries-abilities}   
\end{subfigure}
\hfill
\begin{subfigure}[t]{.48\columnwidth}
    \begin{tikzpicture}
      \begin{axis}[
          height=1.05\columnwidth,
          width=0.95\columnwidth,       
          ymin=0,
          ymax=0.4,          
          xtick={1,2},
          xticklabels={Baseline, \method},
          boxplot/draw direction=y,
          boxplot/box extend=0.3,
      ]
      
      \addplot+[boxplot, mark=*, mark size=1,] table[y=baseline, col sep=comma] {data/harvest_env/allotment/gini_berries_consumed.csv};
      \addplot+[boxplot, mark=*, mark size=1,] table[y=maximin, col sep=comma] {data/harvest_env/allotment/gini_berries_consumed.csv};
    
      \end{axis}
    \end{tikzpicture}
    \caption{Allotment harvest.
    }
    \label{fig:gini-berries-allotment}
\end{subfigure}
\caption{Comparing Gini index of $ag\fsub{well-being}$ and $ag\fsub{resource}$ for $e$. Lower Gini in \method indicates lower inequality.}
\label{fig:gini}
\end{figure}
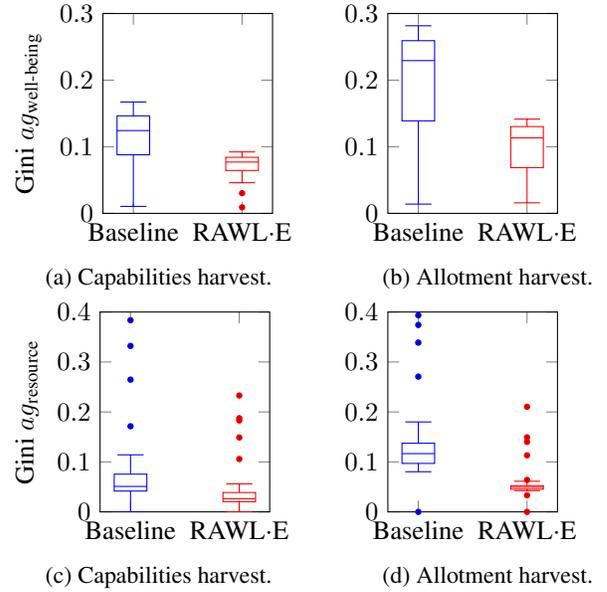

We find that \method societies have lower inequality, indicated by a lower Gini index, in both scenarios. Inequality is especially apparent in the allotment harvest for $ag\fsub{well-being}$, where $\bar{x} = 0.2$ for the baseline society and $\bar{x} = 0.1$ for \method. We reject the null hypothesis corresponding to H\fsub1 as $p<0.01$ for $ag\fsub{well-being}$ and $ag\fsub{resource}$; the effect is large (1.58 for $ag\fsub{well-being}$; 1.32 for $ag\fsub{resource}$).
Figure~\ref{fig:gini} compares Gini index for each society. 

\paragraph{H\fsub2 (minimum experience)}

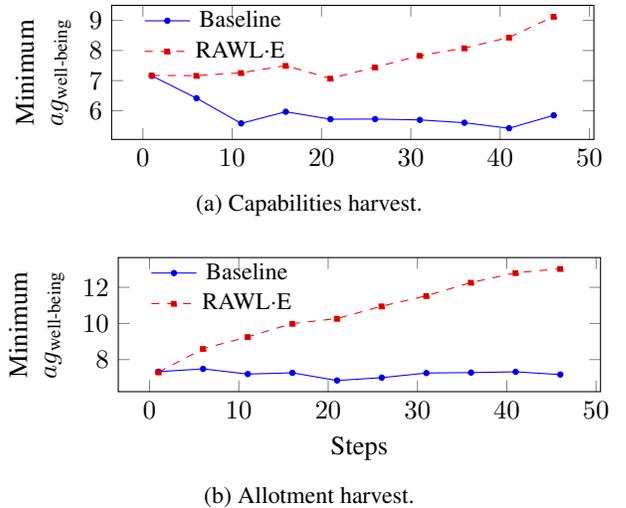
\begin{figure}[!htb]
\begin{subfigure}[t]{\columnwidth}
    \centering
    \begin{tikzpicture}
    \begin{axis}[
    height=0.40\columnwidth,
    width=0.95\columnwidth,
    ylabel style={text width=0.3\columnwidth, align=center},
    ylabel={Minimum\\$ag\fsub{well-being}$},
    legend style={
            at={(0.05,0.50)},
            anchor=south west,
            font=\footnotesize,
            fill=none,
            draw=none,
        },
    ]
    \addplot +[mark size=1, each nth point = {5},style=solid] table [x=day, y=baseline, col sep=comma]
    {data/harvest_env/abilities/min_days_left_to_live.csv};
    \addplot +[mark size=1, each nth point = {5},style=dashed] table [x=day, y=maximin, col sep=comma]
    {data/harvest_env/abilities/min_days_left_to_live.csv};
    \legend{Baseline, \method}
    \end{axis}
    \end{tikzpicture}  
    \caption{Capabilities harvest.
    }
    \label{fig:minimum-well-being-abilities}
\end{subfigure}
\begin{subfigure}[t]{\columnwidth}
    \centering
    \begin{tikzpicture}
    \begin{axis}[
    height=0.40\columnwidth,
    width=0.95\columnwidth,
    xlabel={Steps},
    ylabel style={text width=0.3\columnwidth, align=center},
    ylabel={Minimum\\$ag\fsub{well-being}$},
    legend style={
            at={(0.05,0.50)},
            anchor=south west,
            font=\footnotesize,
            fill=none,
            draw=none,
        },
    ]
    \addplot +[mark size=1, each nth point = {5},style=solid] table [x=day, y=baseline, col sep=comma]
    {data/harvest_env/allotment/min_days_left_to_live.csv};
    \addplot +[mark size=1, each nth point = {5},style=dashed] table [x=day, y=maximin, col sep=comma]
    {data/harvest_env/allotment/min_days_left_to_live.csv};
    \legend{Baseline, \method}
    \end{axis}
    \end{tikzpicture}  
    \caption{Allotment harvest.
    }
    \label{fig:minimum-well-being-allotment}
\end{subfigure}
    \caption{Minimum $ag\fsub{well-being}$ over $t\fsub{max}$ steps summed for $e$, normalised by step frequency. \method yields higher minimum well-being.}
    \label{fig:minimum-well-being}
\end{figure}

\method societies have higher minimum individual experience than baseline agents in both scenarios. 
The largest effect (3.09) is on $ag\fsub{well-being}$ in the allotment harvest, with $\bar{x} = 10.82$ in \method and $\bar{x} = 7.18$ for baseline. For $ag\fsub{well-being}$, we reject the null hypothesis corresponding to H\fsub2 as $p<0.01$.  For  $ag\fsub{resource}$, we cannot reject the null hypothesis in as $p>0.01$. Figures~\ref{fig:minimum-well-being} illustrate results for each society for $ag\fsub{well-being}$. 

\paragraph{H\fsub3 (social welfare)}

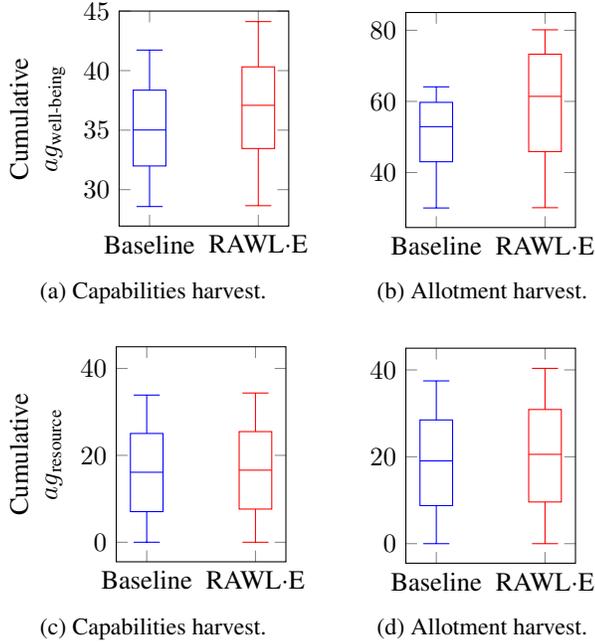
\begin{figure}[!htb]
\centering
\begin{subfigure}[t]{.48\columnwidth}
\centering
    \begin{tikzpicture}
      \begin{axis}[
          height=1.1\columnwidth,
          width=0.95\columnwidth,                   
          ylabel style={text width=0.9\columnwidth, align=center},
          ylabel={Cumulative\\$ag\fsub{well-being}$},
          ymax=45,
          xtick={1,2},
          xticklabels={Baseline, \method},
          boxplot/draw direction=y,
          boxplot/box extend=0.3,
      ]
      
      \addplot+[boxplot, mark=*, mark size=1,] table[y=baseline, col sep=comma] {data/harvest_env/abilities/total_days_left_to_live.csv};
      \addplot+[boxplot, mark=*, mark size=1,] table[y=maximin, col sep=comma] {data/harvest_env/abilities/total_days_left_to_live.csv};
    
      \end{axis}
    \end{tikzpicture}
    \caption{Capabilities harvest.
    }
    \label{fig:cumulative-well-being-abilities}
\end{subfigure}
\hfill
\begin{subfigure}[t]{.48\columnwidth}
\centering
    \begin{tikzpicture}
      \begin{axis}[
          height=1.1\columnwidth,
          width=0.95\columnwidth,          
          ymax=85,
          xtick={1,2},
          xticklabels={Baseline, \method},
          boxplot/draw direction=y,
          boxplot/box extend=0.3,
      ]
      
      \addplot+[boxplot, mark=*, mark size=1,] table[y=baseline, col sep=comma] {data/harvest_env/allotment/total_days_left_to_live.csv};
      \addplot+[boxplot, mark=*, mark size=1,] table[y=maximin, col sep=comma] {data/harvest_env/allotment/total_days_left_to_live.csv};
    
      \end{axis}
    \end{tikzpicture}
    \caption{Allotment harvest.
    }
    \label{fig:cumulative-well-being-allotment}
\end{subfigure}

\begin{subfigure}[t]{.48\columnwidth}
\centering
    \begin{tikzpicture}
      \begin{axis}[
          height=1.1\columnwidth,
          width=0.95\columnwidth,     
          ylabel style={text width=0.9\columnwidth, align=center},
          ylabel={Cumulative\\$ag\fsub{resource}$},
          ymax=45,
          xtick={1,2},
          xticklabels={Baseline, \method},
          boxplot/draw direction=y,
          boxplot/box extend=0.3,
      ]
      
      \addplot+[boxplot, mark=*, mark size=1,] table[y=baseline, col sep=comma] {data/harvest_env/abilities/total_berries_consumed.csv};
      \addplot+[boxplot, mark=*, mark size=1,] table[y=maximin, col sep=comma] {data/harvest_env/abilities/total_berries_consumed.csv};
    
      \end{axis}
    \end{tikzpicture}
    \caption{Capabilities harvest.
    }
    \label{fig:cumulative-berries-abilities}
\end{subfigure}
\hfill
\begin{subfigure}[t]{.48\columnwidth}
\centering
    \begin{tikzpicture}
      \begin{axis}[
          height=1.1\columnwidth,
          width=0.95\columnwidth,          
          ymax=45,
          xtick={1,2},
          xticklabels={Baseline, \method},
          boxplot/draw direction=y,
          boxplot/box extend=0.3,
      ]
      
      \addplot+[boxplot, mark=*, mark size=1,] table[y=baseline, col sep=comma] {data/harvest_env/allotment/total_berries_consumed.csv};
      \addplot+[boxplot, mark=*, mark size=1,] table[y=maximin, col sep=comma] {data/harvest_env/allotment/total_berries_consumed.csv};
    
      \end{axis}
    \end{tikzpicture}
    \caption{Allotment harvest.
    }
    \label{fig:cumulative-berries-allotment}
\end{subfigure}
\caption{Cumulative $ag\fsub{well-being}$ and $ag\fsub{resource}$ of each society over $t\fsub{max}$ steps summed for $e$, normalised by step frequency. Societies of \method agents have higher well-being and cumulative resource consumption.}
\label{fig:cumulative}
\end{figure}

\method yields higher social welfare. 
For $ag\fsub{well-being}$, the allotment harvest yields $\bar{x} = 59.80$ for \method and $\bar{x} = 20.60$ for baseline which has a medium effect (0.64). 
We reject the null hypothesis corresponding to H\fsub3 for $ag\fsub{well-being}$ ($p<0.01$), the difference, however, for $ag\fsub{resource}$ is not significant.
Figure~\ref{fig:cumulative} displays these results.

\paragraph{H\fsub4 (robustness)}

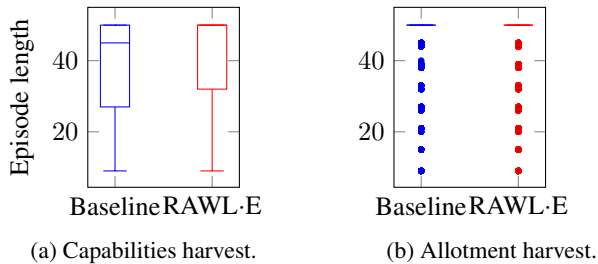
\begin{figure}[!htb]
\begin{subfigure}[t]{.45\columnwidth}
    \begin{tikzpicture}
      \begin{axis}[
          height=1.05\columnwidth,
          width=0.95\columnwidth, 
          ylabel={Episode length},
          ymax=55,
          xtick={1,2},
          xticklabels={Baseline, \method},
          boxplot/draw direction=y,
          boxplot/box extend=0.3,
      ]
      
      \addplot+[boxplot, mark=*, mark size=1,] table[y=day, col sep=comma] {data/harvest_env/abilities/processed_episode_df_baseline.csv};
      \addplot+[boxplot, mark=*, mark size=1,] table[y=day, col sep=comma] {data/harvest_env/abilities/processed_episode_df_maximin.csv};
    
      \end{axis}
    \end{tikzpicture}
    \caption{Capabilities harvest.
    }
    \label{fig:survival-abilities}
\end{subfigure}
\hfill
\begin{subfigure}[t]{.45\columnwidth}
    \begin{tikzpicture}
      \begin{axis}[
          height=1.05\columnwidth,
          width=0.95\columnwidth,    
          ymax=55,
          xtick={1,2},
          xticklabels={Baseline, \method},
          boxplot/draw direction=y,
          boxplot/box extend=0.3,
      ]
      
      \addplot+[boxplot, mark=*, mark size=1,] table[y=day, col sep=comma] {data/harvest_env/allotment/processed_episode_df_baseline.csv};
      \addplot+[boxplot, mark=*, mark size=1,] table[y=day, col sep=comma] {data/harvest_env/allotment/processed_episode_df_maximin.csv};
    
      \end{axis}
    \end{tikzpicture}
    \caption{Allotment harvest.
    }
    \label{fig:survival-allotment}
\end{subfigure}
\caption{Days survived for $e$. Societies of \method agents survive for longer, indicating higher robustness.}
\label{fig:survival}
\end{figure}

\method societies survive longer ($\bar{x} = 48.19$ in allotment) than baseline societies ($\bar{x} = 47.36$ in allotment) indicating higher robustness. We reject the null hypothesis corresponding to H\fsub4 as $p<0.01$; the effect is negligible. 
Figures~\ref{fig:survival-abilities} and \ref{fig:survival-allotment} show results for each society.

\paragraph{Summary of Findings}
Our results support our hypotheses. Our main findings are:  
\begin{enumerate*}[label=(\arabic*)]
\item in a society of \method agents, social welfare is improved, indicated by higher cumulative resource consumption, 
\item  inequality is reduced, indicated by a lower Gini index,
\item  minimum individual experience is higher than the baseline;
the combination of reduced inequality and improved minimum individual experience suggests that \method societies are fairer, and 
\item \method societies survive longer, indicating higher robustness.     
\end{enumerate*}
Together, these results suggest \method agents promote the emergence of norms which improve fairness and social welfare, thereby promoting considerate behaviour, further leading to a more sustainable society.

We observe that results are better (higher fairness, social welfare, and robustness) for \method than baseline in both scenarios.
However, the difference is more apparent in the allotment harvest than capabilities harvest.
We attribute this difference to the fact that in the capabilities harvest agents are in a more confined space than the allotment harvest, and must navigate around one another to reach berries.

\paragraph{Threats to Validity}
Threats arise from the simplicity of our scenarios. While this abstraction limits real-world applicability, our focus is on demonstrating the operationalisation of normative ethics rather than capturing realism. To address this threat, we present our agent architecture decoupled from the environment. Also, using shaped rewards to operationalise ethics offers an adaptable method compatible with various RL algorithms and diverse scenarios.

\section{Discussion and Conclusion}
\label{sec:conclusion}

Developing agents that behave in ways that promote ethical norms is crucial for ethical MAS. Operationalising principles from normative ethics in individual decision making helps address the problem of deriving an ought from an is. 
Our results show that, compared to societies of baseline agents who don't implement normative ethics, \method agents societies have higher social welfare, and are more fair by higher minimum experience and reduced disparity. 

\paragraph{Directions and Key Takeaways}
\label{sec:future-directions}

Applying normative ethics presents challenges, and there is often disagreement on the subject \cite{Moor2006NatureOfMachineEthics}. Conflicts may arise when different principles promote different actions \cite{robinson_moral_2023}. Additionally, the application of a principle may lead to unintuitive outcomes or fail to promote one action over another \cite{guinebert2020MoralTheories}. 
Utilising a variety of principles in reasoning is beneficial to examine scenarios from different perspectives, improving the amplitude of ethical reasoning.
Directions include operationalising a variety of principles, and investigating circumstances in which principles conflict.

We utilise rewards to promote learning ethical behaviour when not all states can be known in advance.
However, modifying rewards combines different objectives in a single numerical scale, allowing implicit comparisons between outcomes \cite{Nashed2023FairnessAS}. 
Directions include combining promotion of ethical behaviour with explicit prevention of unethical outcomes. 


The scenarios we implement are abstracted to demonstrate how the method can be implemented. 
Operationalising normative ethics provides a mechanism to systematically assess the rightness and wrongness of actions in a range of situations \cite{Binns2018FarinessInML}.
Applying our method to more complex and real world scenarios, with a range of different RL algorithms, is another direction for future work.


\paragraph{Reproducibility}
Our codebase is publicly available \cite{AAAI25-Rawle-Codebase}.
Appendices~\ref{appx:experimental-setups}--\ref{appx:simulation-results} provide additional details, including computing infrastructure, parameter selection, a complete list of environmental rewards, further descriptions of metrics, a complete set of emerged norms, and additional details on simulation results.

\section{Acknowledgments}
JW thanks EPSRC Doctoral Training Partnership Grant No. EP/W524414/1 for the support. 
NA thanks UKRI EPSRC Grant No. EP/Y028392/1: AI for Collective Intelligence (AI4CI) for the support.

\bibliography{Nirav,Jess}

\newpage

\appendix

\clearpage
\section{Details of Experimental Setups}
\label{appx:experimental-setups}

This section provides details about the computing infrastructure and hyperparameter selection.

\subsection{Computing Infrastructure}
\label{appx:computing-infrastructure}

We conducted the simulation experiments on a workstation with Intel Xeon Processor W-2245 (8C 3.9 GHz), 256GB RAM, and Nvidia RTA A6000 48GB GPU.

\subsection{Hyperparameter Selection}
\label{appx:hyperparameter-selection}

Table~\ref{tbl:simulation-parameter-selection} lists the simulation parameters and range of values tried per parameter. Results are consistent across the range of values tried, with societies of \method agents having higher social welfare, fairness, and robustness than societies of baseline agents.

\begin{table}[!hb]
    \centering
    \caption{Parameters for simulation experiments.}
    \label{tbl:simulation-parameter-selection}
    \begin{tabular}{p{2cm}p{1.3cm}cS}
        \toprule 
         {Description} & {Parameter} & {Range Tried} & {Final Value}
        \\\midrule
        Capabilities grid size & $o\fsub{capabilities} \times p\fsub{capabilities}$ & \{$4 \times 4$, $8 \times 4$\}& {$8 \times 4$}
        \\\midrule
        Allotment grid size & $o\fsub{allotment} \times p\fsub{allotment}$ & \{$8 \times 4$, $16 \times 4$\} & {$16 \times 4$}
        \\\midrule
        Number of agents & $k$ & \{2, 4\} & 4
        \\\midrule
        Initial number of berries & $b\fsub{initial}$ & \{8, 12, 16\} & 12
        \\\midrule
        Initial health of agent & $h\fsub{initial}$ & \{5.0, 10.0\} & 5.0
        \\\midrule
        Health gain from eating berry & $h\fsub{gain}$ & \{0.1, 1.0\} & 0.1
        \\\midrule
        Health decay & $h\fsub{decay}$ & \{-0.01, -0.1\} & -0.01
        \\\midrule
        Minimum health to throw & $h\fsub{throw}$ & \{0.5, 0.6, 1.0\} & 0.6
        \\\midrule
        Number of episodes & $e$ & \{1000, 2000\} & 2000
        \\\midrule
        Maximum steps in episode & $t\fsub{max}$ & \{20, 50\} & 50
        \\\bottomrule
    \end{tabular}
\end{table}

Table~\ref{tbl:decision-parameter-selection} lists the interaction module parameters and range of values tried per parameter. We select these parameters empirically, with reference to literature \cite{Bengio2012deepArchitectures}.

\begin{table*}[!htb]
    \centering
    \caption{Parameters of the Interaction Module.}
    \label{tbl:decision-parameter-selection}
    \begin{tabular}{lclSl}
        \toprule 
         {Description} & {Parameter} & {Range Tried} & {Final Value} & {Criterion}
        \\\midrule
        Batch size & $B$ & \{32, 64, 128\} & 64 & Training time
        \\\midrule
        Iteration for updating weights of target network & $C$ & \{1000, 100, 50\} & 50 & Test performance
        \\\midrule
        Probability of exploration & $\epsilon$ & {0.9--0.0} & 0.0 & Test performance
        \\\midrule
        Learning rate & $\alpha$ & \{0.01, 0.001, 0.0001\} & 0.0001 & Test performance
        \\\midrule
        Number of hidden units & ${Hn}$ & \{32, 64, 128\} & 128 & Test performance
        \\\midrule
        Number of hidden layers & ${Hl}$ & {1--3} & 2 & Test performance
        \\\bottomrule
    \end{tabular}                           
\end{table*}

\section{Environmental Rewards}
\label{appx:rewards}

To avoid obvious results by giving \method agents additional rewards, we normalise rewards between baseline and \method agents such that \method agents receive lower raw rewards. This allows for fairer comparison between societies. Table~\ref{tbl:complete-rewards} displays the complete list of rewards received by baseline and \method agents.

\begin{table}[!htb]
    \sisetup{round-mode=places,
             round-precision=2,
             table-column-width=3em 
             }

    \centering
    \caption{Rewards received by an agent. Rewards are normalised between baseline and \method agents to avoid obvious results by giving \method agents more rewards.}
    \label{tbl:complete-rewards}
    \begin{tabular}{@{}p{5.4cm} S S}
        \toprule 
         {Action} & {Baseline} & {\method}
        \\\midrule
        Survive episode & 1.0 & 1.0
        \\
        Eat berry & 1.0 & 0.8
        \\
        Forage where berry is & 1.0 & 0.8
        \\
        Throw berry & 0.5 & 0.5
        \\
        Try to eat without berries & -0.2 & -0.1
        \\
        Try to throw without berries & -0.2 & -0.1
        \\
        Try to throw without sufficient health & -0.2 & -0.1
        \\
        Try to throw without recipient & -0.2 & -0.1
        \\
        Die & -1.0 & -1.0
        \\
        Improve minimum experience & 0.0 & 0.4
        \\
        No difference to minimum experience & 0.0 & 0.0
        \\
        Did not improve minimum experience & 0.0 & -0.4
        \\\bottomrule
    \end{tabular}
\end{table}

\section{Metrics}
\label{appx:metrics}

Here, we provide further details about the metrics used to evaluate societies of \method and baseline agents.

To assess the fairness of a society, we compute M\fsub1 (inequality) and M\fsub2 (minimum experience).

\begin{description}[nosep,leftmargin=0em]
    \item[M\fsub1 (inequality)] Examining the inequality of a society to assess fairness is supported by the principle of egalitarianism, which states that disparity amongst members should be minimised \cite{Murukannaiah-AAMAS20-BlueSky}. We use the Gini index for inequality as it is well studied and has been used previously in MAS \cite{Endriss2013Inequality}. 
        
    \item[M\fsub2 (minimum experience)] Examining the minimum individual experience to assess fairness is justified by Rawlsian ethics, which argues that those who benefit the least should be prioritised \cite{Rawls2001JusticeFairnessRestatement}.
\end{description}

The fairest society will have the lowest inequality and highest minimum individual experience. 
However, the notion of fairness is abstract and achieving perfect fairness is challenging, if not impossible \cite{Dignum2021FairnessMyth}. 
We thus aim for satisfactory outcomes that promote equitable systems, which have a higher goal of fairness, but might not be perfect.

To assess the society's sustainability, we compute the metrics M\fsub3~(social welfare) and M\fsub4~(robustness).
\begin{description}[nosep,leftmargin=0em]
    \item[M\fsub3 (social welfare)] Measuring social welfare (how much society as a whole gains \cite{Mashayekhi+TAAS22-Cha}) is supported by the principle of utilitarianism, which states that ethical actions are those which maximise utility \cite{Ong2024RideScheduling}. 
    
    \item[M\fsub4 (robustness of society)] Robustness relates to the degree a society is sensitive to exogenous influence, exhibited as the ability to resist and withstand adversity \cite{Munoz2022ResilienceRobustness}.
\end{description}

\section{Results for Emerged Norms}

\subsection{Results for Emerged Cooperative Norms}
\label{appx:cooperative-norms}

A norm is emerged when it is adopted by over 90\% of the population. In societies of \method agents, the cooperative norms which emerge have higher fitness and are used more frequently, indicated by higher numerosity. Table~\ref{tbl:norms-summary} summarises the results for emerged cooperative norms.

\begin{table*}[!htb]
    \centering
    \caption{Comparing fitness and numerosity of cooperative norms of baseline and \method societies in capabilities harvest and allotment harvest scenarios. Grey highlight indicates best results with significance at $p<0.01$.}
    \label{tbl:norms-summary}
    
    \begin{tabular}{llSSSSS}
        \toprule
                    \multirow{2}{*}{Scenario} & \multirow{2}{*}{Metrics} & \multicolumn{2}{c}{Mean $\bar{x}$} 
                    & \multicolumn{2}{c}{Standard deviation $\sigma$}  
                    & \multicolumn{1}{c}{\multirow{2}{*}{Cohen's d}} \\
                    \cmidrule(lr){3-4}
                    \cmidrule(lr){5-6}
                    & & {Baseline} & {\method} & {Baseline} & {\method} &  \\\midrule
    \multirow{2}{*}{\parbox{1.5cm}{Capabilities Harvest}}
       & {Fitness}
       & 25.61 & \cellcolor{gray!20}57.62 & 41.28 & 83.44 & 0.3 \\
       & {Numerosity} 
       & 14.61 & \cellcolor{gray!20}26.58 & 16.4 & 20.54 & 0.25 \\\midrule
    \multirow{2}{*}{\parbox{1.5cm}{Allotment Harvest}}
       & {Fitness}
       & 46.5 & \cellcolor{gray!20}64.34 & 84.59 & 92.51 & 0.26 \\
       & {Numerosity} 
       & 26.32 & \cellcolor{gray!20}41.94 & 24.57 & 32.28 & 0.26 \\ \bottomrule       
    \end{tabular}
\end{table*}

\subsection{System of Emerged Norms}
\label{appx:emerged-norms}

\begin{figure*}[!htb]
\begin{subfigure}[t]{\textwidth}
    \centering
        \begin{tikzpicture}[
            treenode/.style = {align=center, inner sep=0pt, text centered},
            level 1/.style = {level distance = 2.8cm, sibling distance = 2cm},
            level 2/.style = {level distance = 2.8cm, sibling distance = 1.5cm},
            level 3/.style = {level distance = 2.8cm, sibling distance = 1cm},
            level 4/.style = {level distance = 2.8cm, sibling distance = 0.8cm},
            edge from parent/.style={draw, edge from parent path={(\tikzparentnode.east) -- +(0.25cm,0) |- (\tikzchildnode)}},
            grow'=right
        ]
        
        \node[treenode][yshift=2cm]{medium health}
            child {node[yshift=-0.5cm] {no berries}
                child {node {\textbf{move}}}
            }
            child {node[yshift=-1cm] {medium berries}
                child {node {medium days}
                    child {node {medium days}
                        child {node[yshift=0.5cm] {low days}
                            child {node {\textbf{eat}}}
                            child {node {\textcolor{blue}{\textbf{throw}}}}
                        }
                        child {node {medium days}
                            child {node {\textbf{eat}}}
                            child {node {\textcolor{blue}{\textbf{throw}}}}
                        }
                    }
                    child {node {low days}
                        child {node {\textcolor{blue}{\textbf{throw}}}}
                    }
                    child {node {high days}
                        child {node {\textcolor{blue}{\textbf{throw}}}}
                    }
                }
                child {node {low days}
                    child {node {\textbf{eat}}}
                }
            }
            child {node[yshift=-0.5cm] {low health}
                child {node {\textbf{move}}}
            };
        
        \node[treenode][yshift=-1.5cm] {low health}
            child {node[yshift=-0.5cm] {no berries}
                child {node {\textbf{move}}}
            }
            child {node {medium berries}
                child {node {medium days}
                    child {node[yshift=0.5cm] {medium days}
                        child {node {\textbf{eat}}}
                        child {node {\textbf{move}}}
                    }
                    child {node {low days}
                        child {node {\textbf{eat}}}
                        child {node {\textbf{move}}}
                    }
                }
                child {node {low days}
                    child {node {\textbf{eat}}}
                    child {node {high days}
                        child {node {\textbf{move}}}
                    }
                }
            };
        
        \node[treenode][yshift=-4.5cm] {high health}
            child {node[yshift=-0.5cm] {no berries}
                child {node {\textbf{move}}}
            }
            child {node[yshift=0.5cm] {medium berries}
                child {node {\textbf{eat}}}
            };
        \end{tikzpicture}
    \caption{Baseline norms}
    \label{fig:norms-allotment-baseline}
\end{subfigure}

\begin{subfigure}[t]{\textwidth}
    \centering
        \begin{tikzpicture}[
            treenode/.style = {align=center, inner sep=0pt, text centered},
            level 1/.style = {level distance = 2.8cm, sibling distance = 2cm},
            level 2/.style = {level distance = 2.8cm, sibling distance = 1.5cm},
            level 3/.style = {level distance = 2.8cm, sibling distance = 1cm},
            level 4/.style = {level distance = 2.8cm, sibling distance = 0.8cm},
            edge from parent/.style={draw, edge from parent path={(\tikzparentnode.east) -- +(0.25cm,0) |- (\tikzchildnode)}},
            grow'=right
        ]
        
        \node[treenode][yshift=2cm]{medium health}
            child {node[yshift=-0.5cm] {no berries}
                child {node {\textbf{move}}}
            }
            child {node {medium berries}
                child{node {medium days}
                    child{node {\textbf{eat}}
                    }
                    child{node {medium days}
                        child{node {low days}
                            child{node{\textcolor{blue}{\textbf{throw}}}
                            }
                            child{node{medium days}
                                child{node{\textbf{move}}
                                }
                            }
                        }
                    }
                    child{node {low days}
                        child{node {medium days}
                            child{node{low days}
                                child{node{\textbf{move}}
                                }
                            }
                        }
                    }
                }
                child{node[yshift=-1cm] {low days}
                    child{node{\textcolor{blue}{\textbf{throw}}}
                    }
                    child{node{medium days}
                        child{node{low days}
                            child{node{\textbf{move}}
                            }
                        }
                    }
                    child{node{low days}
                        child{node{\textbf{move}}
                        }
                    }
                }
            };
        
        \node[treenode][yshift=-2cm] {low health}
            child {node[yshift=-0.5cm] {no berries}
                child {node {\textbf{move}}
                }
            }
            child {node[yshift=0.5cm] {medium berries}
                child{node[yshift=-0.5cm]{\textbf{eat}}
                }
                child{node[yshift=0.5cm]{low days}
                    child{node{low days}
                        child{node{medium days}
                            child{node{\textbf{move}}
                            }
                        }
                    }
                }
            };
            
        \node[treenode, yshift=-4cm] {high health}
            child {node[yshift=-0.5cm] {no berries}
                child {node {\textbf{move}}}
            }
            child {node {medium berries}
                child {node {high days}
                    child {node {\textcolor{blue}{\textbf{throw}}}}
                    child {node {high days}
                        child {node {\textbf{eat}}}
                    }
                }
                child {node {medium days}
                    child {node {\textcolor{blue}{\textbf{throw}}}}
                }
            };
        \end{tikzpicture}
    \caption{\method norms}
    \label{fig:norms-allotment-method}
    \end{subfigure}
    \caption{$\mathcal{N}$ for allotment harvest over $e\fsub{epochs}$. \textcolor{blue}{blue} highlights cooperative norms. In societies of \method agents, cooperative norms which emerge are more generalised than in baseline societies.}
    \label{fig:norms-allotment}
\end{figure*}
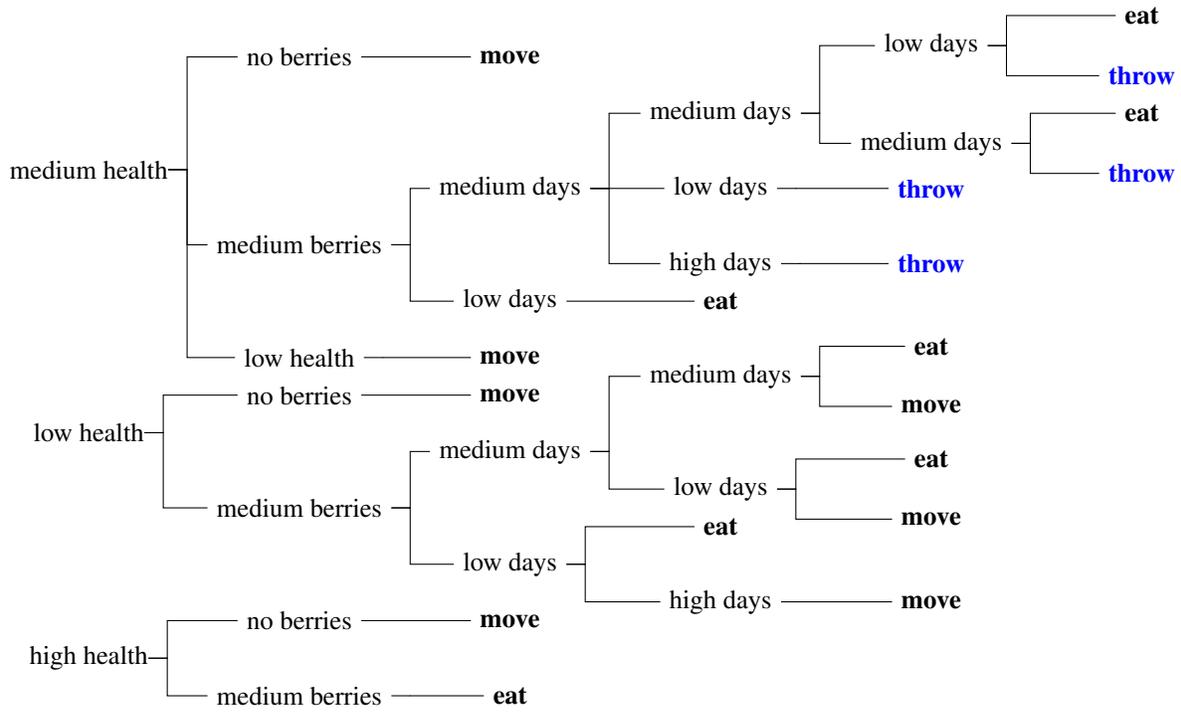
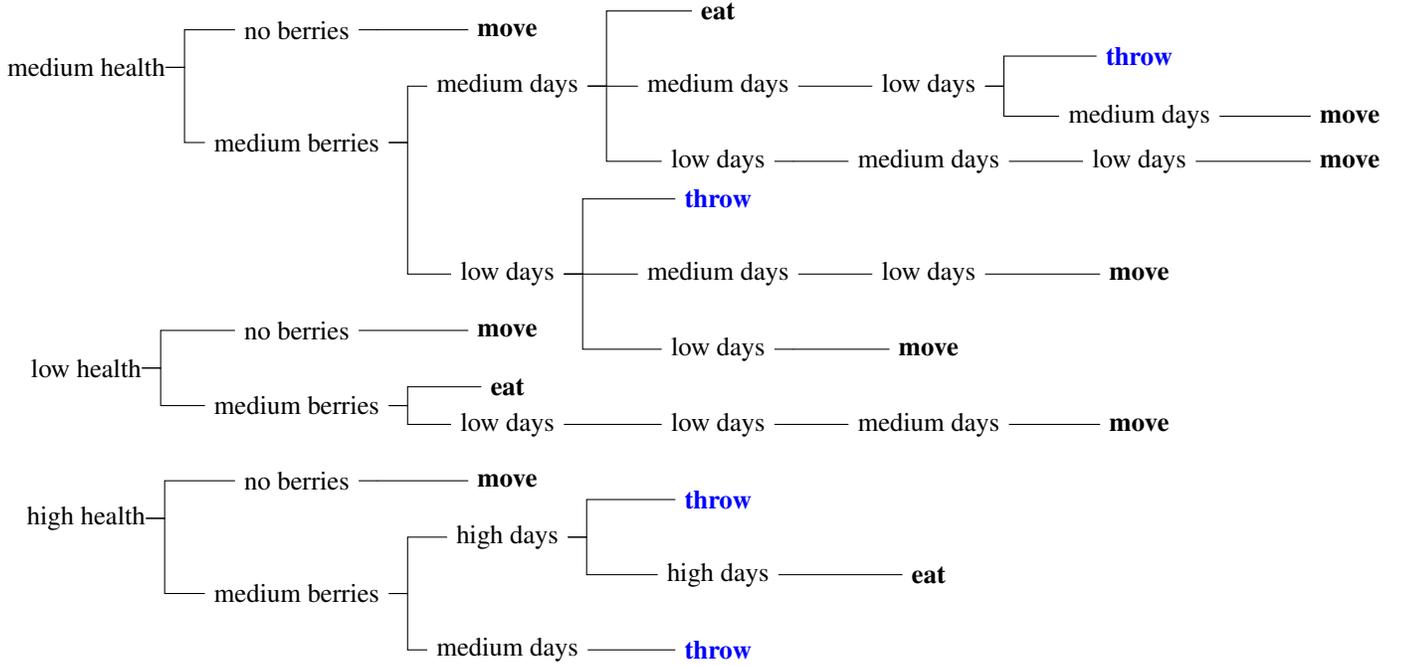

Figures~\ref{fig:norms-capabilities} and ~\ref{fig:norms-allotment} display the system of emerged norms $\mathcal{N}$ in baseline and \method societies. To obtain $\mathcal{N}$, we run $e=2000$ episodes for $t\fsub{max}=50$ steps and track the norms which emerge in each episode. We combine $\mathcal{N}$ for $e$ to obtain the list of all norms which emerge. Norms with ``throw" consequent are cooperative, as throwing is an act of agents helping one another. To distill specific norms into generalised rules, we aggregate antecedent conditions which produce the same outcome. For example, all instances of the condition ``no berries" result in the consequent ``move". Therefore, specific norms are aggregated to the generalised rule of:

\begin{lstlisting}[numbers=none]
    IF <no berries> THEN <move>
\end{lstlisting}

In both scenarios, we observe that in societies of \method agents cooperative norms which emerge are more generalised than cooperative norms emerging in societies of baseline agents. For example, in Figure~\ref{fig:norms-capabilities-method} a general norm emerges:

\begin{lstlisting}[numbers=none]
    IF <high health> THEN <throw>
\end{lstlisting}

In contrast, the cooperative norms which emerge in baseline societies in Figure~\ref{fig:norms-capabilities-baseline} are more specialised than in \method societies. This indicates that cooperative norms in \method societies cover a wider range of circumstances.

\begin{figure*}[!htb]
\begin{subfigure}[t]{\textwidth}
    \centering
        \begin{tikzpicture}[
            treenode/.style = {align=center, inner sep=0pt, text centered},
            level 1/.style = {level distance = 3cm, sibling distance = 3cm},
            level 2/.style = {level distance = 3cm, sibling distance = 2cm},
            level 3/.style = {level distance = 3cm, sibling distance = 1.5cm},
            edge from parent/.style={draw, edge from parent path={(\tikzparentnode.east) -- +(0.25cm,0) |- (\tikzchildnode)}},
            grow'=right
        ]
        
        \node[treenode][yshift=2cm]{medium health}
            child {node[yshift=-1cm] {no berries}
                child {node {\textbf{move}}}
            }
            child {node {medium berries}
                child {node {medium days}
                    child {node {medium days}
                        child {node[yshift=-0.5cm] {\textbf{eat}}}
                        child {node {\textcolor{blue}{\textbf{throw}}}}
                    }
                    child {node {low days}
                        child {node {\textcolor{blue}{\textbf{throw}}}}
                    }
                    child {node {high days}
                        child {node {\textcolor{blue}{\textbf{throw}}}}
                    }
                }
                child {node {low days}
                    child {node {\textcolor{blue}{\textbf{throw}}}}
                }
            };
        
        \node[treenode][yshift=-2cm] {low health}
            child {node[yshift=-1cm] {no berries}
                child {node {\textbf{move}}}
            }
            child {node[yshift=1cm] {medium berries}
                child {node {\textbf{eat}}}
            };
        \end{tikzpicture}
    \caption{Baseline norms}
    \label{fig:norms-capabilities-baseline}
\end{subfigure}

\begin{subfigure}[t]{\textwidth}
    \centering
        \begin{tikzpicture}[
            treenode/.style = {align=center, inner sep=0pt, text centered},
            level 1/.style = {level distance = 3cm, sibling distance = 2cm},
            level 2/.style = {level distance = 3cm, sibling distance = 1.5cm},
            edge from parent/.style={draw, edge from parent path={(\tikzparentnode.east) -- +(0.25cm,0) |- (\tikzchildnode)}},
            grow'=right
            ]
            
            \node[treenode][yshift=3cm]{medium health}
                child {node[yshift=-0.5cm] {no berries}
                    child {node {\textbf{move}}}
                }
                child {node[yshift=0.3cm] {medium berries}
                    child {node {low days}
                        child {node {\textbf{eat}}}
                    }
                    child {node {medium days}
                        child {node {medium days}
                            child {node {\textcolor{blue}{\textbf{throw}}}}
                            child {node {\textbf{eat}}}
                        }
                        child {node {low days}
                            child {node {\textbf{eat}}}
                        }
                    }
                };
            
            \node[treenode]{low health}
                child {node[yshift=-0.5cm] {no berries}
                    child {node {\textbf{move}}}
                }
                child {node[yshift=0.5cm] {medium berries}
                    child {node {\textbf{eat}}}
                };
            
            \node[treenode][yshift=-1.5cm] {high health}
                child {node[yshift=-0.5cm] {no berries}
                    child {node {\textbf{move}}}
                }
                child {node[yshift=0.5cm] {\textcolor{blue}{\textbf{throw}}}};
                \end{tikzpicture}
    \caption{\method norms}
    \label{fig:norms-capabilities-method}
    \end{subfigure}
    \caption{$\mathcal{N}$ for capabilities harvest over $e\fsub{epochs}$. \textcolor{blue}{blue} highlights cooperative norms. In societies of \method agents, more generalised cooperative norms emerge than in baseline societies. For example, in Figure~\ref{fig:norms-capabilities-method} the general norm of ``IF high health THEN throw" emerges.}
    \label{fig:norms-capabilities}
\end{figure*}
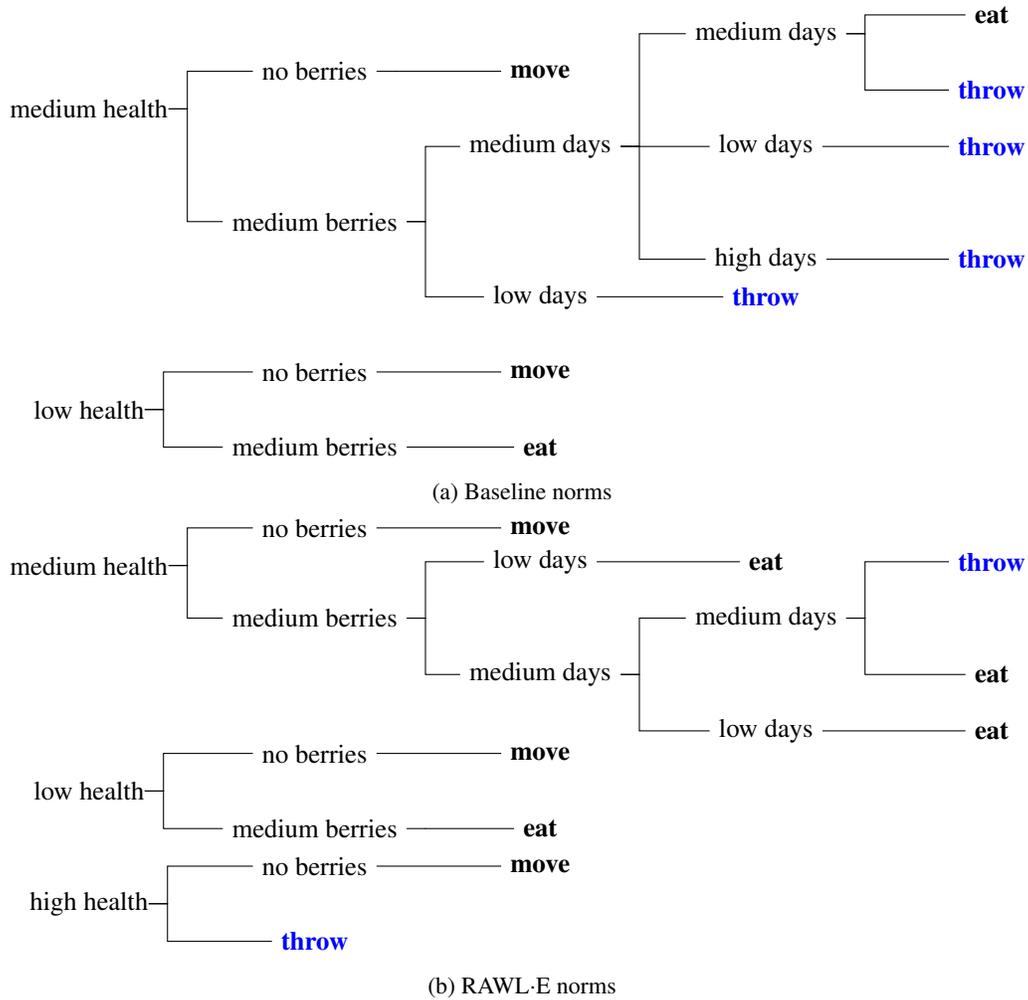

\section{Simulation Results}
\label{appx:simulation-results}

Table~\ref{tbl:results-total} provides additional details of the simulation results. 

We observe that societies of \method agents have significantly lower inequality than societies of baseline agents for both $ag\fsub{well-being}$ and $ag\fsub{resource}$. The effect is medium to large in both scenarios: 1.58 for $ag\fsub{well-being}$ and 0.63 for $ag\fsub{resource}$ in the capabilities harvest; 1.58 for $ag\fsub{well-being}$ and 1.32 for $ag\fsub{resource}$ in the allotment harvest. 

For minimum experience, societies of \method agents show significantly higher results than baseline societies for $ag\fsub{well-being}$ in both scenarios with a large effect, however, the minimum experience is not significantly different for $ag\fsub{resource}$. 

Social welfare is significantly higher for $ag\fsub{well-being}$ in societies of \method agents than baseline societies in the allotment harvest with a medium effect of 0.64. The difference in social welfare is not significant for $ag\fsub{resource}$. 

Further, in both scenarios, societies of \method agents are more robust than baseline societies; however, the effect is negligible (0.18 in capabilities harvest and 0.11 in allotment harvest).

\begin{table*}[!h]
    \centering
    \caption{Comparing $ag\fsub{resource}$, inequality, minimum experience, and robustness of baseline and \method societies in capabilities harvest and allotment harvest scenarios. Grey highlight indicates best results with significance at $p<0.01$.}
    \label{tbl:results-total}
    
    \begin{tabular}{lllSSSSS}
        \toprule
                    \multirow{2}{*}{Scenario} & \multirow{2}{*}{Metrics} & \multirow{2}{*}{Variable} & \multicolumn{2}{c}{Mean $\bar{x}$} 
                    & \multicolumn{2}{c}{Standard deviation $\sigma$}  
                    & \multicolumn{1}{c}{\multirow{2}{*}{Cohen's d}} \\
                    \cmidrule(lr){4-5}
                    \cmidrule(lr){6-7}
                    & & & {Baseline} & {\method} & {Baseline} & {\method} &  \\\midrule
    \multirow{8}{*}{\parbox{1.5cm}{Capabilities Harvest}} 
       & \multirow{2}{*}{M\fsub1 Inequality} 
       & {$ag\fsub{well-being}$} & 0.12 & \cellcolor{gray!20}0.07 & 0.04 & 0.02 & 1.58\revisionst{0.58} \\
       & & {$ag\fsub{resource}$} & 0.09 & \cellcolor{gray!20}0.04 & 0.1 & 0.05 & 0.63\revisionst{0.53} \\
       & \multirow{2}{*}{M\fsub2 Minimum experience} 
       & {$ag\fsub{well-being}$} & 5.81 & \cellcolor{gray!20}7.69 & 0.36 & 0.62 & 3.71\revisionst{0.85} \\
       & & {$ag\fsub{resource}$} & 3.62 & 3.98 & 2.41 & 2.52 & 0.15\revisionst{0.077} \\
       & \multirow{2}{*}{M\fsub3 Social welfare }
       & {$ag\fsub{well-being}$} & 35.25 & 37.05 & 3.62 & 4.39 & 0.45\revisionst{0.22} \\
       & & {$ag\fsub{resource}$} & 16.52 & 16.96 & 10.41 & 10.43 & 0.04\revisionst{0.017} \\
       & {M\fsub4 Robustness} &  & 38.07 & \cellcolor{gray!20}40.5 & 14.15 & 13.48 & 0.18\revisionst{0.096} \\\midrule
    \multirow{8}{*}{\parbox{1.5cm}{Allotment Harvest}}
       & \multirow{2}{*}{M\fsub1 Inequality} 
       & {$ag\fsub{well-being}$} & 0.2 & \cellcolor{gray!20}0.1 & 0.08 & 0.04 & 1.58\revisionst{0.6} \\
       & & {$ag\fsub{resource}$} & 0.14 & \cellcolor{gray!20}0.06 & 0.08 & 0.03 & 1.32\revisionst{0.73} \\
       & \multirow{2}{*}{M\fsub2 Minimum experience} 
       & {$ag\fsub{well-being}$} & 7.18 & \cellcolor{gray!20}10.82 & 0.23 & 1.65 & 3.09\revisionst{0.85} \\ 
       & & {$ag\fsub{resource}$} & 3.79 & 4.5 & 2.43 & 2.73 & 0.27\revisionst{0.13} \\
       & \multirow{2}{*}{M\fsub3 Social welfare} 
       & {$ag\fsub{well-being}$} & 51.5 & \cellcolor{gray!20}59.8 & 9.9 & 15.41 & 0.64\revisionst{0.29} \\
       & & {$ag\fsub{resource}$} & 18.94 & 20.6 & 11.35 & 12.28 & 0.14\revisionst{0.07} \\
       & {M\fsub4 Robustness} & & 47.36 & \cellcolor{gray!20}48.19 & 7.67 & 6.94 & 0.11\revisionst{0.089} \\ \bottomrule       
    \end{tabular}
\end{table*}

\end{document}